\documentclass[12pt,letterpaper]{article}
\usepackage[a4paper, total={7in, 10in}]{geometry}

\usepackage{graphicx}
\usepackage{helvet}
\usepackage{authblk}
\usepackage{hyperref}
\usepackage{amsmath} 
\usepackage{amssymb} 
\usepackage[super,comma,sort&compress]  
   {natbib}
\usepackage[right]{lineno} \linenumbers
\usepackage{subcaption}
\usepackage{url} 

\makeatletter
\renewcommand{\maketitle}{\bgroup\setlength{\parindent}{0pt}
\begin{flushleft}
  \textbf{\@title}
  
  \@author
\end{flushleft}\egroup}
\makeatother

\title{Collaborative AI in Sentiment Analysis: System Architecture, Data Prediction and Deployment Strategies}
\date{}

\author[1]{Chaofeng Zhang}
\author[2]{Jia Hou}
\author[3]{Xueting Tan}
\author[3]{Gaolei Li}
\author[4]{Caijuan Chen}

\affil[1]{Advanced Institute of Industrail Technology, Tokyo, Japan}
\affil[2]{Soochow University, No.333 Ganjiang Road, Suzhou, Jiangsu, China}
\affil[3]{Shanghai Jiaotong University, No.800 Dong Chuan Roa, Shanghai, China}
\affil[4]{National Insituite of Informaticst, Chiyoda City, Tokyo, Japan}

\begin{document}

\maketitle

\section*{SUMMARY}

The advancement of large language model (LLM) based artificial intelligence technologies has been a game-changer, particularly in sentiment analysis. This progress has enabled a shift from highly specialized research environments to practical, widespread applications within the industry. However, integrating diverse AI models for processing complex multimodal data and the associated high costs of feature extraction presents significant challenges. Motivated by the marketing oriented software development needs, our study introduces a collaborative AI framework designed to efficiently distribute and resolve tasks across various AI systems to address these issues. Initially, we elucidate the key solutions derived from our development process, highlighting the role of generative AI models like \emph{chatgpt}, \emph{google gemini} in simplifying intricate sentiment analysis tasks into manageable, phased objectives. Furthermore, we present a detailed case study utilizing our collaborative AI system in edge and cloud, showcasing its effectiveness in analyzing sentiments across diverse online media channels.

\section*{KEYWORDS}

%%%  Include up to 10 keywords, separated by commas. 
%%%  Keywords entered in EM are not carried over; only 
%%%  keywords included in the main text will be used 
%%%  in the final article metadata.    

Collaborative artificial intelligence, sentiment analysis architecture, multimodal data processing, edge computing deployment, algorithmic prompt enhancement

\section*{INTRODUCTION}

Sentiment analysis technologies have evolved from being confined to research laboratories to attracting substantial interest from both the market and social organizations. This shift coincides with the progressive maturation of general AI technologies \cite{Yang2020Sentiment}. Demonstrating stable performance, these technologies adeptly manage the voluminous and diverse information generated by millions of online users daily \cite{Jelodar2020Deep}. Data from various sources, including official media, blogs, forums, and social networks, undergoes preliminary processing before sentiment analysis algorithms analyze it to deduce insights into positive and negative sentiments. In our highly interconnected world, the outcomes of such analyses significantly impact both workflows and corporate decision-making processes. For instance 		\cite{Poria2023Beneath}, before launching new products, company decision-makers can utilize sentiment analysis to gauge public opinion during holidays, thereby forecasting potential online engagement, market reception, and sales projections.

\begin{figure}
	\centerline{\includegraphics[width=30.5pc]{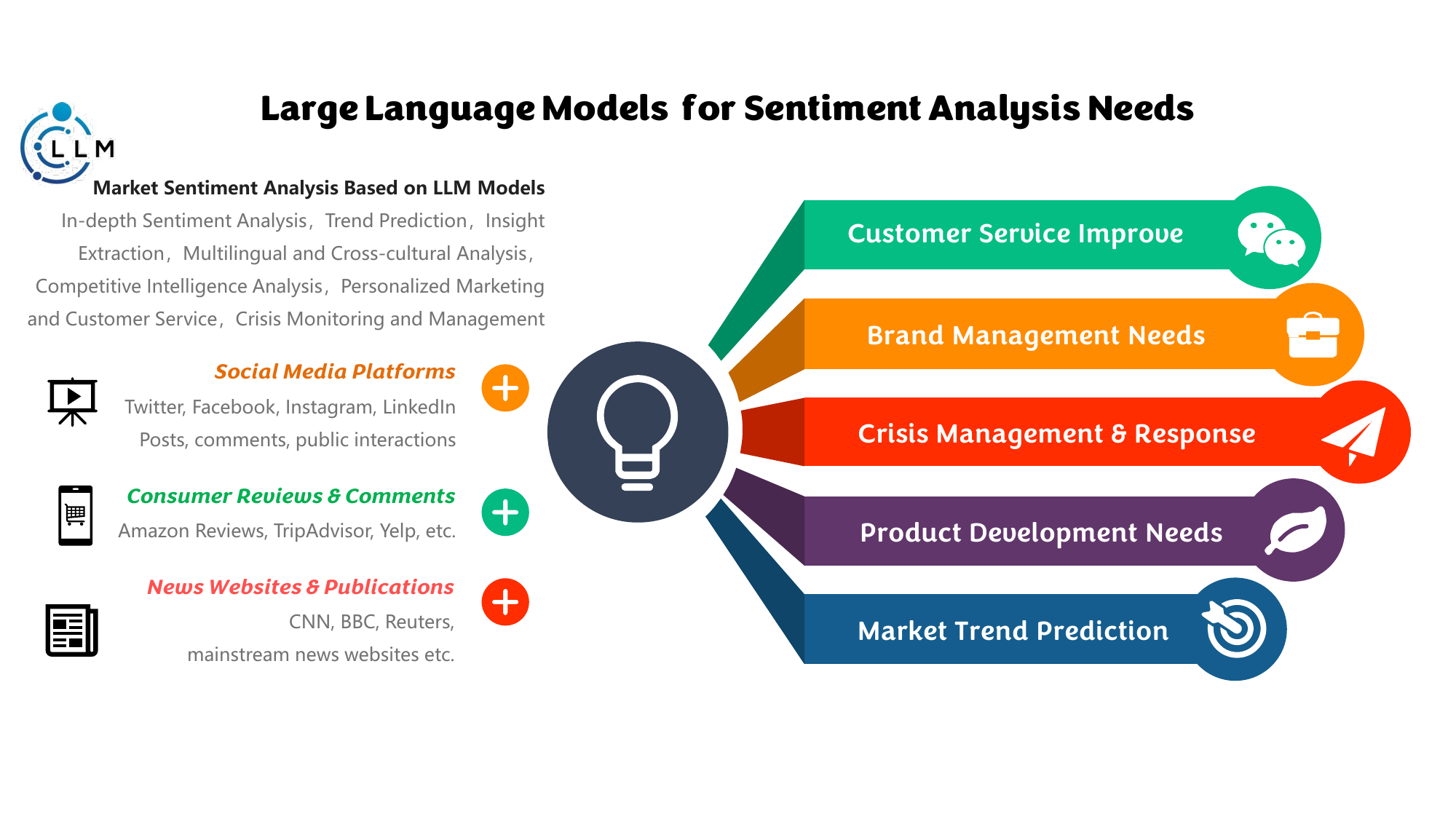}}
	\caption{Sentiment analysis is highly valued in today's market. People utilize diverse media channels to interpret multimodal data for the analysis of market acceptance of products or services.}
	\label{fig:marketneeds}
\end{figure}

To foster more objective sentiment analysis, as shown in Fig.~\ref{fig:marketneeds}, people employ web scraping, crowdsourcing, and APIs to amass vast quantities of data from online media \cite{Imran2020Cross}. The multimodal nature of this data, encompassing video, audio, text, and numerical formats, presents considerable challenges and escalates the cost of feature extraction \cite{Galassi2021Attention}. Nevertheless, the progressively stable advancements in AI technologies have markedly enhanced the accuracy of sentiment analysis. Massive multimodal data sets are processed and transformed into features or sentiment scores using advanced neural network models, such as Convolutional Neural Networks (CNNs), Recurrent Neural Networks (RNNs), Graph Neural Networks (GNNs), and Transformer models. By synthesizing insights from various data modalities, sentiment analysis systems achieve a level of accuracy unattainable through single-modal analysis. These systems are now capable of sophisticated functions like sarcasm detection, negation handling, and spam identification.

Before the advent of prominent large models like ChatGPT and Gemini, the application of various AI technologies, despite ensuring judgment accuracy, still posed numerous issues.
Despite these advancements, integrating varied AI technologies to ensure accuracy presents several challenges 	\cite{Misra2021IoT}. For market analysis or AI-based consulting, as depicted in Fig. 1, different AI models are required for distinct data modalities: CNNs for images and videos, Transformer models for textual data, and Bidirectional Encoder Representations from Transformers (BERT) for analyzing interpersonal dynamics. This necessitates complex, difficult-to-integrate systems. Furthermore, specific sentiment analysis scenarios often demand unique training datasets and finely-tuned neural networks \cite{kaddour2023challenges}, which significantly elevate costs regarding data, intellectual capital, computational resources, and time, thereby hindering rapid adaptation to new scenarios. An additional challenge lies in the interpretability of neural network computations \cite{Ghuribi2020Unsupervised}, which creates a gap between sentiment scores and the original data, complicating the justification and traceability of results. 
Fortunately, the emergence of generative artificial intelligence (GAI) potential \cite{Ooi2020potential}, exemplified by ChatGPT-4o, offers potential solutions. Its core large language model (LLM) demonstrates high flexibility and versatility in processing natural language tasks, further broadening the applicable scenarios and impact of sentiment analysis.

However, LLM-based sentiment analysis systems encounter several challenges in executing complex tasks 	\cite{zhang2023building}. For example, while LLMs such as ChatGPT are adept at multitasking, their performance in intricate multitasking scenarios, particularly those involving logical reasoning and result integration, needs to be revised. Additionally, LLMs often need help comprehending nuanced human prompts primarily designed for general scenarios 	\cite{Wang2023Chain}. Translating each individual's unique thought process into natural language and ensuring its accurate interpretation by LLM models is a significant hurdle. To overcome these limitations, collaborative AI has emerged as a promising solution. This approach involves assigning specific roles and methodologies to different AI systems and distributing tasks among various agents for systematic and probabilistic resolution. Such collaboration facilitates meticulous execution and reasoning at different stages \cite{ziems2023large}, culminating in a comprehensive analytical report for the users.

To overcome the problem of multinational data and high cost in conventional sentiment analysis, this paper aims to delve into the potential of collaborative AI in specific social applications, mainly focusing on interactive AI development that influences governmental and business decisions, such as the analysis of trending social topics. We discuss common issues with LLMs in practical development, including the incomplete absorption of prompt meanings. A detailed application paradigm is outlined, demonstrating how to efficiently design a collaborative AI system architecture, encompassing system definition, architecture design, operational mechanisms, and implementation examples. Ultimately, we aim to present a one-click software solution that offers advantages over traditional approaches in collaborative AI sentiment analysis architecture, delivering accurate predictions and recommendations in the final report to users.

By examining the practical development needs of projects, this study innovatively provides an inclusive development paradigm for diverse tasks. It explores the feasibility of collaborative AI in the field of public sentiment analysis and offers the following contributions for subsequent large-scale sentiment analysis and development:
\begin{enumerate}
	\item \textbf{Broadening Collaborative AI Paradigm}: The study redefines the objectives of tasks and feedback according to diverse public sentiment analysis requirements, employing different collaborative AI modules to expand the scope of customized tasks.
	\item \textbf{Automate Sentiment Analysis Architecture}: A collaborative architecture for integrated analysis of historical, current, and future sentiment data is established, achieving a broader and more accurate sentiment analysis than independent LLMs or traditional neural networks.
	\item \textbf{Develop Large Model Language Prompt Algorithm}: By summarizing numerous practical developments, algorithm-based prompts have been developed to enhance the stability of sentiment report outputs.
	\item \textbf{Conduct Feasibility Validation}: Validation  of the collaborative sentiment AI system is conducted across multiple dimensions, including runtime, sentiment analysis accuracy, and local deployment feasibility, thereby enhancing the capability for differentiated deployment of collaborative sentiment analysis AI.
\end{enumerate}

\section*{Emerging Trends in Collaborative AIs}
\begin{figure*}
	\centerline{\includegraphics[width=40.5pc]{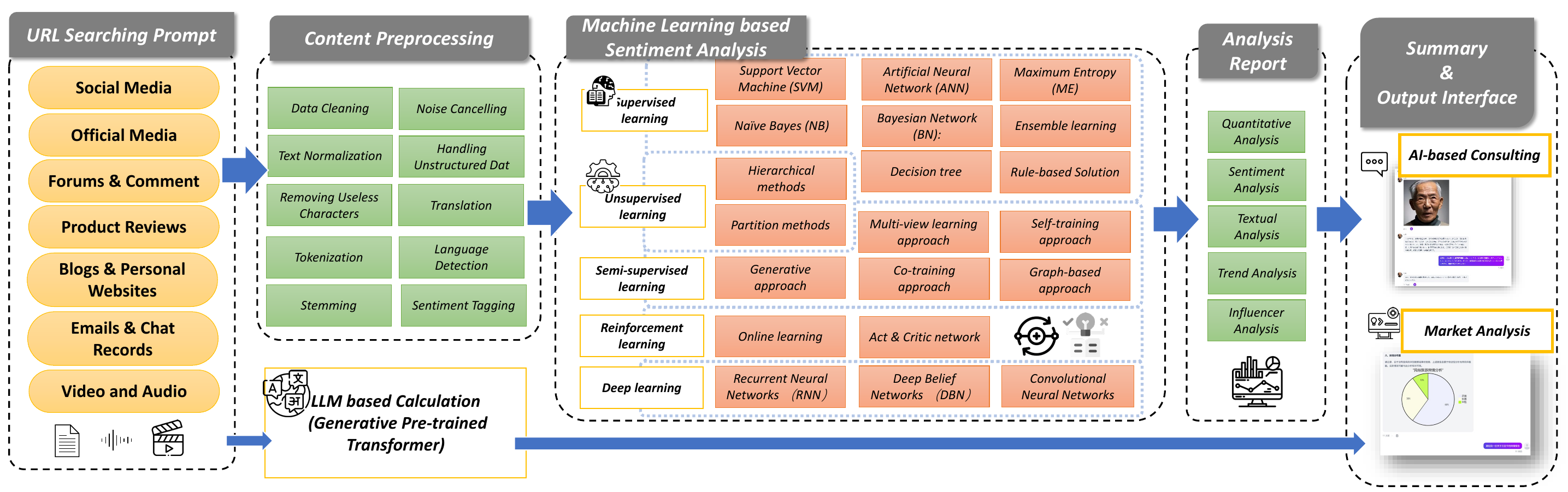}}
	\caption{Cutting-edge sentiment analysis solutions employing diverse AI technologies to analyze text, images, and sound, then computing sentiment-related scores to realize public opinion analysis or AI-based consulting.}
	\label{fig:SentimentAIApplication}
\end{figure*}

Fig.~\ref{fig:SentimentAIApplication} summarizes the general methodology for generating public sentiment reports in an environment dominated by online media. Data from various types of online media can be processed and summarized by both non-LLM and LLM models, ultimately providing refined analytical results for decision-makers. Collaborative AI demonstrates superior capabilities in handling multi-threaded tasks or tasks with complex logic compared to a single LLM. Consequently, the emergence of various collaborative AI frameworks offers significant inspiration for the development of sentiment analysis systems:

%Collaborative AI has proven superior to individual Large Language Models (LLMs) in managing tasks characterized by multi-threaded or complex logical processes. Recent developments in various collaborative AI frameworks have offered substantial insights into the future of artificial intelligence.

\textbf{Simulate Social Interactions}:
Researchers like Joon Sung Park et al. \cite{park2023generative} have successfully simulated credible human behavior by integrating large language models with computational interactive agents. They developed modules for daily scheduling, long-term memory, reflection, and reasoning to enable LLM-based agents to mimic human behavioral patterns. The simulation involving 25 agents revealed that memory prompts are insufficient for effective social interaction and human-like existence. Reflection and reasoning are crucial in collaborative AI, necessitating human-designed phased thinking objectives for agents. This approach facilitates the integration of all thought processes, supporting the agents' rational actions and even simulating emergent group behaviors.

\textbf{Enhance Performance}:
Collaborative AI demonstrates strong adaptability in optimizing tasks that require complex logic. For instance, in software development, which typically involves design, coding, testing, and documentation phases, Chen Qian et al. \cite{qian2023communicative} proposed specialized roles for programmers, reviewers, and testing engineers. Their system, CHATDEV, conducts multiple rounds of code construction and testing, ultimately producing user manuals with fewer flaws than those generated by general GAI models. This collaborative approach has been shown to address potential flaws and errors efficiently and cost-effectively, which is particularly important as user requirements often lead to inaccuracies in LLM-generated code.

\textbf{Facilitate Human-Machine Collaboration}:
The potential of human-machine collaboration, another facet of collaborative AI, is also noteworthy. In response to human behaviors, researchers like Ran Gong et al. \cite{gong2023mindagent} incorporated LLMs (such as GPT-4, Claude, and LLaMA) into the scenario game CUISINEWORLD, achieving complex task scheduling. Their multi-agent planning framework, comprising scene, memory, and action modules, inputs rules, scene information, and roles into LLMs to generate corresponding action plans. This approach yields more reasonable suggestions by understanding human intentions, facilitating efficient operations under human guidance, and significantly reducing communication costs.

\textbf{Navigate Adversarial Interaction}:
In studies conducted by Kai Xiong \cite{xiong2023examining}, debates among groups of agents with varying capabilities revealed that LLMs exhibit different levels of self-consistency and compromise based on their performance. Utilizing superior LLMs to lead collaborations among other LLMs is a feasible strategy. The adversarial interaction and compromise processes are crucial steps for collaborative AI to produce final results.

\section*{LLM based Co-AI Architecture}

\subsection*{Definition of Analysis Task }

The demand for hotspot sentiment analysis primarily emanates from administrative departments, enterprises, commercial entities, and financial institutions. For example, such analysis is utilized to discern public sentiment towards educational policies or specific events, aiding in assessing public response to policy initiatives. In this scenario, we assume that the user organization has already integrated general artificial intelligence applications (such as AI customer service assistants and AI query databases) into their office systems, enabling access to the sentiment analysis assistant via the same interface. Users articulate their analysis requirements through a general chatbot's dialogue box, specifying \textbf{keywords, timestamps, and preference parameters}. The architecture of the hotspot sentiment analysis system gathers information from servers or the internet, compiling it into a report tailored to specific user needs, presented in a human-computer interaction dialogue box. In contrast to traditional data tables and digitized summaries, LLMs provide comprehensive reports ranging from overviews to detailed analyses in both textual and pictorial formats, making them more accessible to decision-makers and laypersons alike.

\textbf{System Objective}: When users seek to understand the social perception of a specific topic, they can directly query the chatbot with requests like \emph{"Provide me with a sentiment analysis report on the Halloween Holiday"} or \emph{"Predict the emotional trend of the Halloween Holiday from October 1, 2019"}. Based on the query, the system assigns a crawler AI to collect information from designated portals and integrates this data into a database. After the database AI processes the timestamp data, a report-writing AI generates the report. Given the questionable reliability of LLM-generated information, all arguments collected by AI are cited with sources (with an option to hide or display URLs), ensuring that the final report generated by the collaborative AI maintains interpretability and contextual logic at every stage, thereby presenting a compelling and persuasive conclusion for users to consider.

\textbf{Analysis Objects: }When users do not specify, the system defaults to gathering information from official and social media sources, considering all mined content for the final report. Regional media preferences and habits dictate the default mainstream media channels for sentiment data collection, such as Yahoo in Japan, Google in English-speaking regions, and Baidu in China. Popular platforms like Line in East Asia, Twitter in English-speaking regions, and Weibo in Chinese-speaking regions are considered for social media. Users can specify intelligence sources or their weightings for a more locally resonant sentiment analysis report.

\subsection*{ System Composition}
As shown in Fig.~\ref{fig:Structure}, the \textbf{structure of Co-AI sentiment analysis} system includes the following modules:

\begin{figure*}
	\centerline{\includegraphics[width=42.5pc]{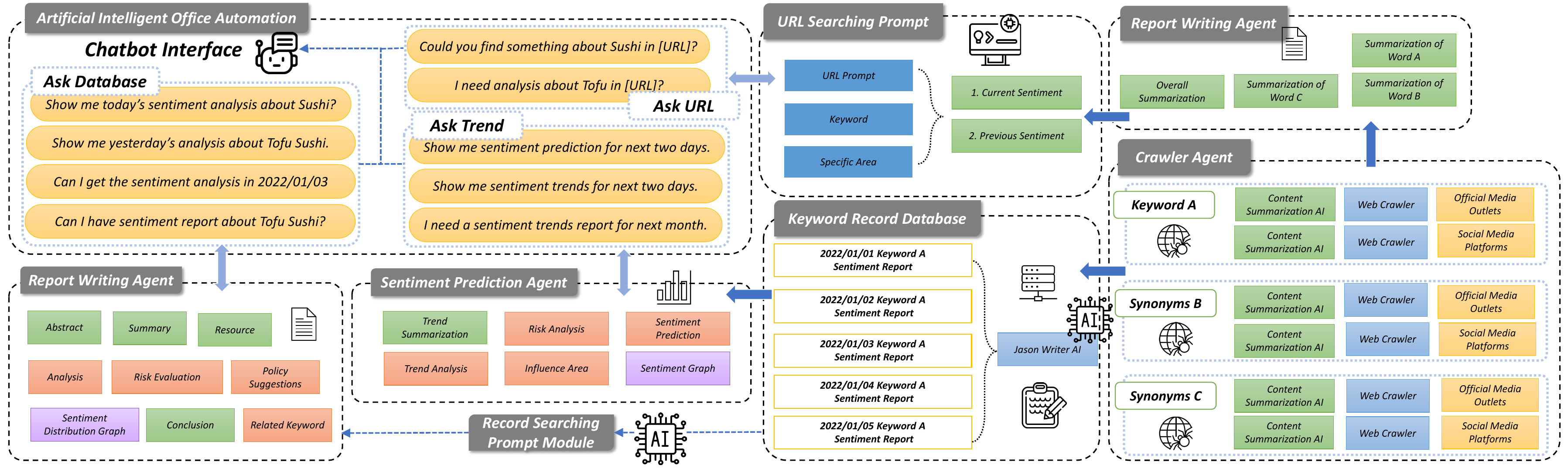}}
	\caption{The proposed architecture of the Co-AI sentiment analysis system, featuring components such as a chatbot interface, report writing agent, record databases, and a crawler agent, designed to compile sentiment analysis reports on \emph{past, present,} and \emph{future} events.}
	\label{fig:Structure}
\end{figure*}

\emph{Chatbot:} This component, capable of invoking LLM APIs and accessing internal databases, serves as the interface for users to request sentiment analysis. It supports three primary report types: database record generation, sentiment trend inquiries, and URL-specific sentiment analysis.

\emph{Report Writing Agent:} Utilizing LLM, this agent generates comprehensive sentiment analysis reports with warnings and policy suggestions. The default setting provides a daily public sentiment overview, including an introduction, overall assessment, and visual representations like pie charts of emotional sentiments and high-frequency associated words.

\emph{Prediction Agent:} This agent forecasts future emotional trends, incorporating risk assessments and impact scopes based on historical data, thereby enhancing decision-making accuracy.

\emph{URL Searching Prompt and Report Writing:} This module fetches data directly from specified websites, with a timestamp attention mechanism to distinguish the temporal relevance of emotional evaluations.

\emph{Record Database:} This module archives daily emotional records of specific keywords in a JSON format for efficient retrieval and integration with LLMs, optimizing API call costs and processing time.

\emph{Crawler Agent:} It collects web content related to specified keywords from default official and social media sources for summarization into emotional reports by various AIs.

\textbf{Interaction of Modules:}  
The Artificial Intelligence Office Automation system serves as the user interface for our sentiment analysis framework, providing comprehensive reports on the \textbf{past, present,} and \textbf{future} sentiments associated with specific keywords. Report Writing and Prediction Agents rely on database records to generate custom reports. Through its program, the Crawler Agent accumulates data in the database or feeds it directly to the URL-based Searching Agent. A notable aspect is using the LLM model's API for generating phased summaries, JSON formatted data, or predictions. Distinctive prompts trigger varied tasks, culminating in an integrated output by a single agent. This collaborative synergy of AI modules enables the generation of holistic sentiment analysis reports.

\subsection*{Generate Analysis Report}
This section delineates the system's objective of producing a comprehensive sentiment analysis report. This report encompasses detailed summaries, in-depth analyses, and specific word-related data. It is structured to ensure a reading time of no more than 10 minutes, enabling decision-makers to assimilate essential information rapidly. A standardized report format has been adopted to minimize reading time. This format leverages Language Model (LLM) prompt-based outputs to create specific sections. The composition of a typical report (termed the Report Writing Agent) incorporates three key segments: \textbf{Summarizing Outputs} (comprising an introduction, a summary, and a conclusion), \textbf{Analytical Outputs} (encompassing cause analysis, risk assessment, policy recommendations, and associated word analysis), and \textbf{Graphical Outputs} (translating data into visual representations). The LLM generates preliminary content for these eight sections based on structured prompts. Subsequently, this content is meticulously reorganized to align with the predefined standard text format before being archived. The foundational texts for these standards are derived from outputs generated by our crawler agent. These texts are then progressively augmented by various AI systems, each responsible for a specific function, culminating in the assembly of the final report. The tasks executed by individual AI components rely on pre-existing data and do not entail autonomous generation, ensuring that the accuracy of the analytical outputs remains both explainable and traceable. Predictive reports and those with constrained search parameters are designed to simplify and partially substitute the prediction and summarization processes while retaining the steps' interpretability and verifiability.

\begin{figure}
	\centerline{\includegraphics[width=34.5pc]{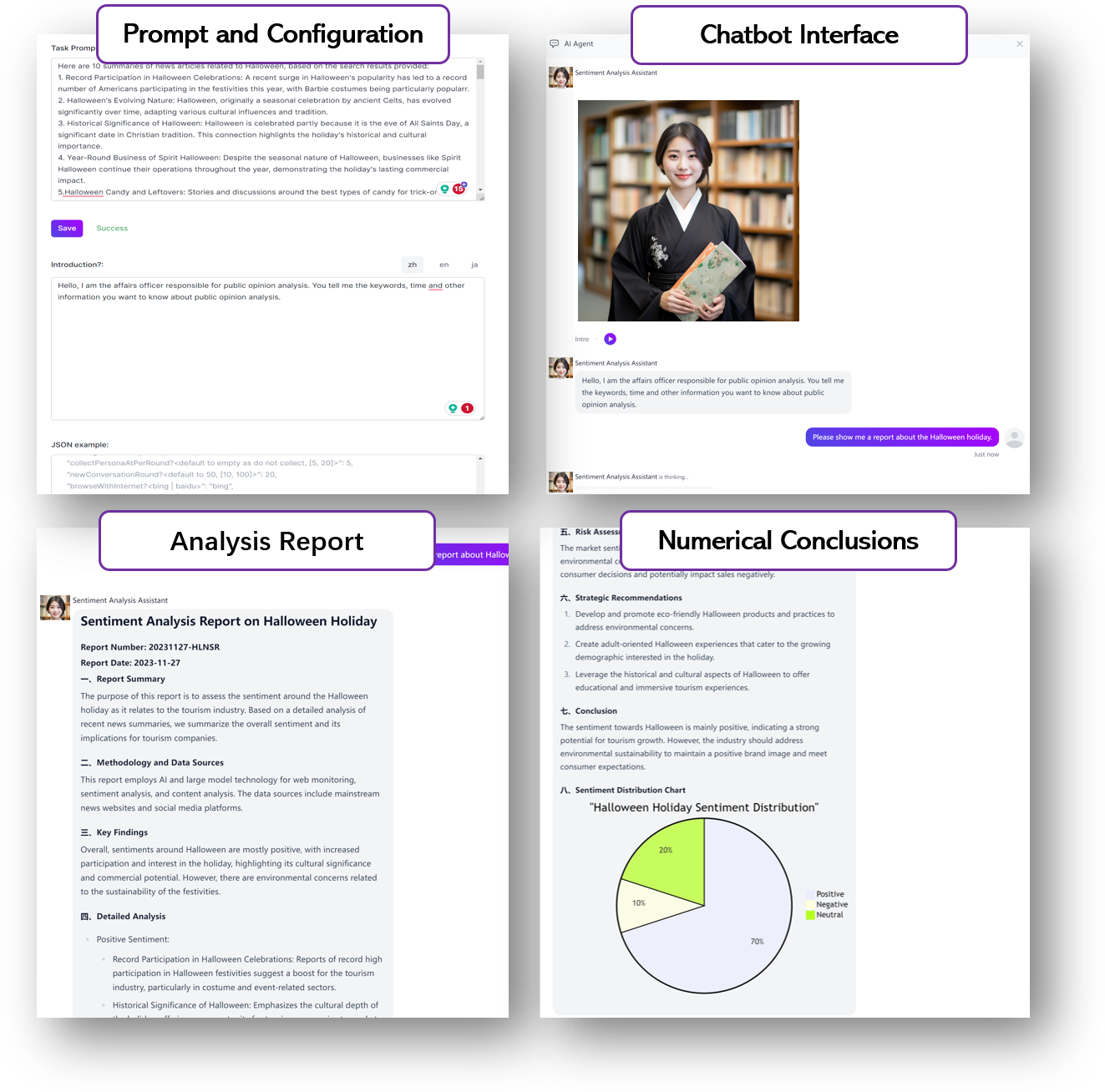}}
	\caption{Interface for Configuring and Reviewing Sentiment Analysis Reports. }
	\label{fig:UI}
\end{figure}

The implementation of this collaborative system is illustrated in Fig.\ref{fig:UI}. The interactive interface is bifurcated into settings and chatbot modules. Within the settings module, users can meticulously customize the report's presentation style and data sources. The chatbot module encompasses a suite of human-machine interaction functionalities, including voice input and output, video interactions, and chart displays. These modules can seamlessly integrate with other chatbot systems deployed within an enterprise or organization, such as online Q$\&$A and query systems.

\section*{ Key Contributions in Collaborative AI Development}
\subsection*{Prompt Design}
\textbf{Agent Role Prompt} forms the cornerstone of each independent AI's functioning within the Co-AI framework, critically influencing the Quality of Experience (QoE) in analysis services. Given the subjective nature of user inputs, the primary aim of prompt design is to facilitate LLMs' comprehension of requirements within these subjective narratives. Recognizing that LLM inputs and outputs are integral to natural language-based human interaction, we systematically incorporate \textbf{Habermas's Theory} of Communicative Action to standardize AI's social behavior logic. 
%Prompt settings encompass Dramaturgical Action for shaping AI's unique perspectives and roles, Goal-Oriented Action for defining AI's objective tasks, Normatively Regulated Action for outlining adherence rules and risk mitigation, and Communicative Action for harmonizing these aspects, where the interplay of subjective input and LLM output accuracy determines user expectation fulfillment. 
For instance, a system prompt for a sentiment analysis AI might be structured as follows:
\begin{enumerate}
	\item \emph{You are a public opinion analysis expert with strong research and analytical capabilities and a deep understanding of the market sales field and related events.}
	\item \emph{Conduct sentiment and content analysis to provide positive and negative evaluations of public opinion.}
	\item \emph{Interpret public opinion information in the correct context and pay attention to data collection and analysis details to ensure information accuracy.}
	\item \emph{You must produce reports based on user input requirements.}
\end{enumerate}
These align with (a) Dramaturgical, (b) Goal-Oriented, (c) Normatively Regulated, and (d) Communicative Actions, respectively. Extensive testing revealed that this prompt structure effectively mitigates the risk of unanticipated responses.

\begin{figure}
	\centerline{\includegraphics[width=32.5pc]{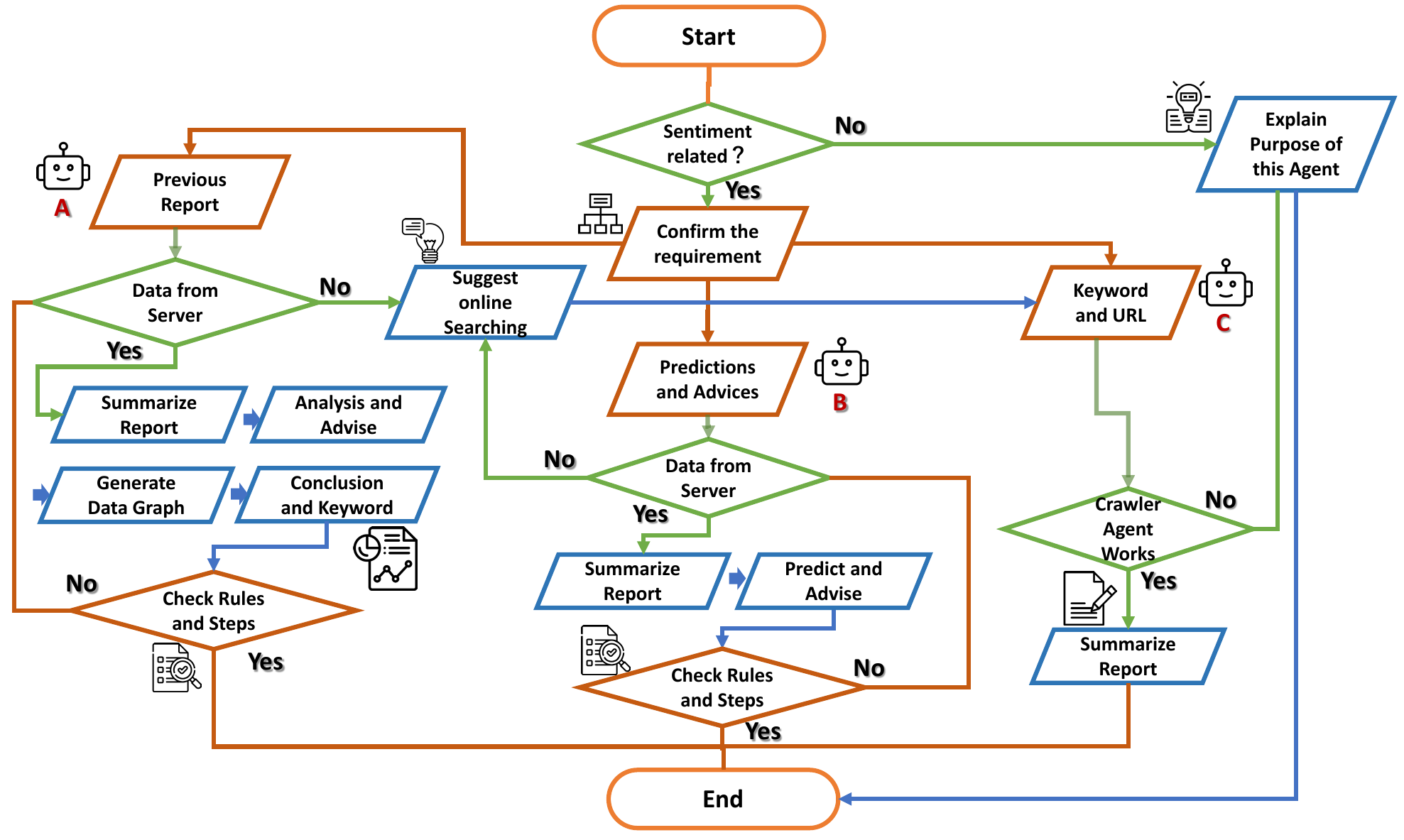}}
	\caption{Development of LLM Thinking Algorithm Flowchart. Adhering to this logic yields more acceptable sentiment analysis reports.}
\end{figure}

\textbf{Thinking Logic}
is pivotal for generating standard reports. Present LLMs cannot process varied report details simultaneously for desired outcomes. To overcome this, we develop a thinking logic algorithm (illustrated in Fig.~4) guiding LLMs in sequential thought processing. For example, in the report writing agent module, prompts include:
\begin{enumerate} % 设置标签格式为小写字母
	\item \emph{Please summarize reports on keywords and synonyms.}
	\item \emph{Compare, dissect, and analyze these emotional pieces of information, summarizing public opinion, and give an overall evaluation of the current sentiment state.}
	\item \emph{Analyze and output the causes and potential impacts of the current sentiment state.}
	\item \emph{Provide risk warnings and improvement suggestions based on the current public sentiment.}
	\item \emph{Separately analyze the causes and trends of the current emotions, adding corresponding titles at the beginning of paragraphs.}
	\item \emph{Complete conclusions, associated words, and other elements.}
	\item \emph{Invoke the image generation API to produce emotion distribution graphs.}
	\item \emph{Final Check: Please think step by step, do not output the thought process, and do not explain the output logic.}
\end{enumerate}

Through extensive testing, we confirm that this algorithmic approach significantly enhances LLM output consistency, addressing the shortcomings of traditional prompts that might present conflicting logic or overlook certain requirements. This method reinforces the LLM's multi-step logic accuracy.

\section*{ Case Study with Numerical Analysis}
In this section, the proposed collaborative AI system leverages highly acclaimed LLM developed by OpenAI and Google to validate the feasibility of both cloud-based and local deployments. Then, we focus on the topic of "food delivery" and analyze public sentiment trends across various platforms, including online media (\emph{Bing News}, \emph{Google News}), search engines (\emph{Google Search}), and social media (\emph{Twitter}, \emph{Yahoo Hot}). Then, we employ \textbf{TextBlob} to conduct general sentiment analysis on the collected data.

\subsection*{LLMs on Cloud and Edge}
\begin{figure}
	\centering
	\begin{subfigure}[b]{0.5\textwidth}
		\centering
		\includegraphics[width=\textwidth]{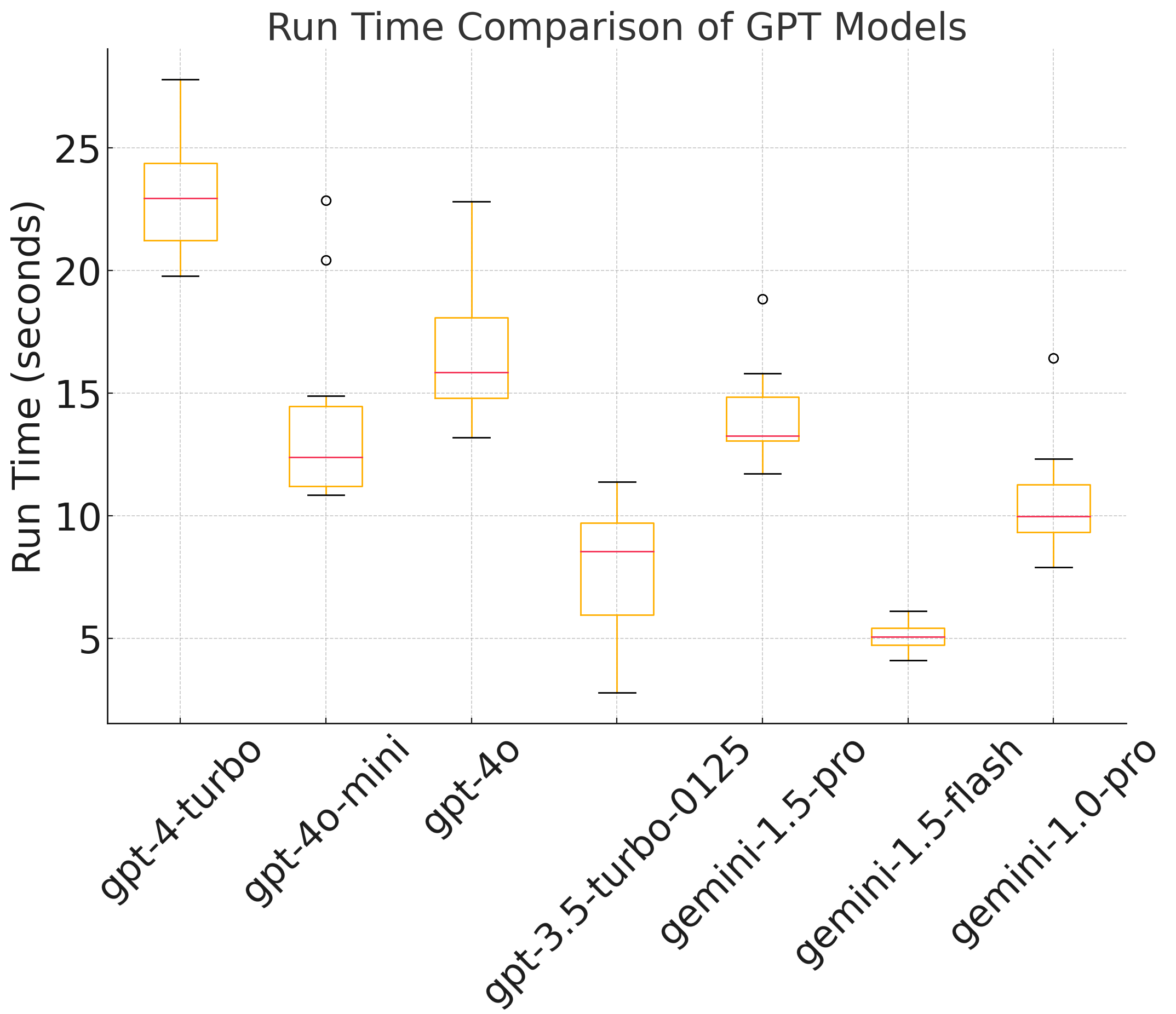}
		\caption{Runtime comparison of different LLMs.}
		\label{fig:RunTimeCloudComparison}
	\end{subfigure}
	\hfill
	\begin{subfigure}[b]{0.48\textwidth}
		\centering
		\includegraphics[width=\textwidth]{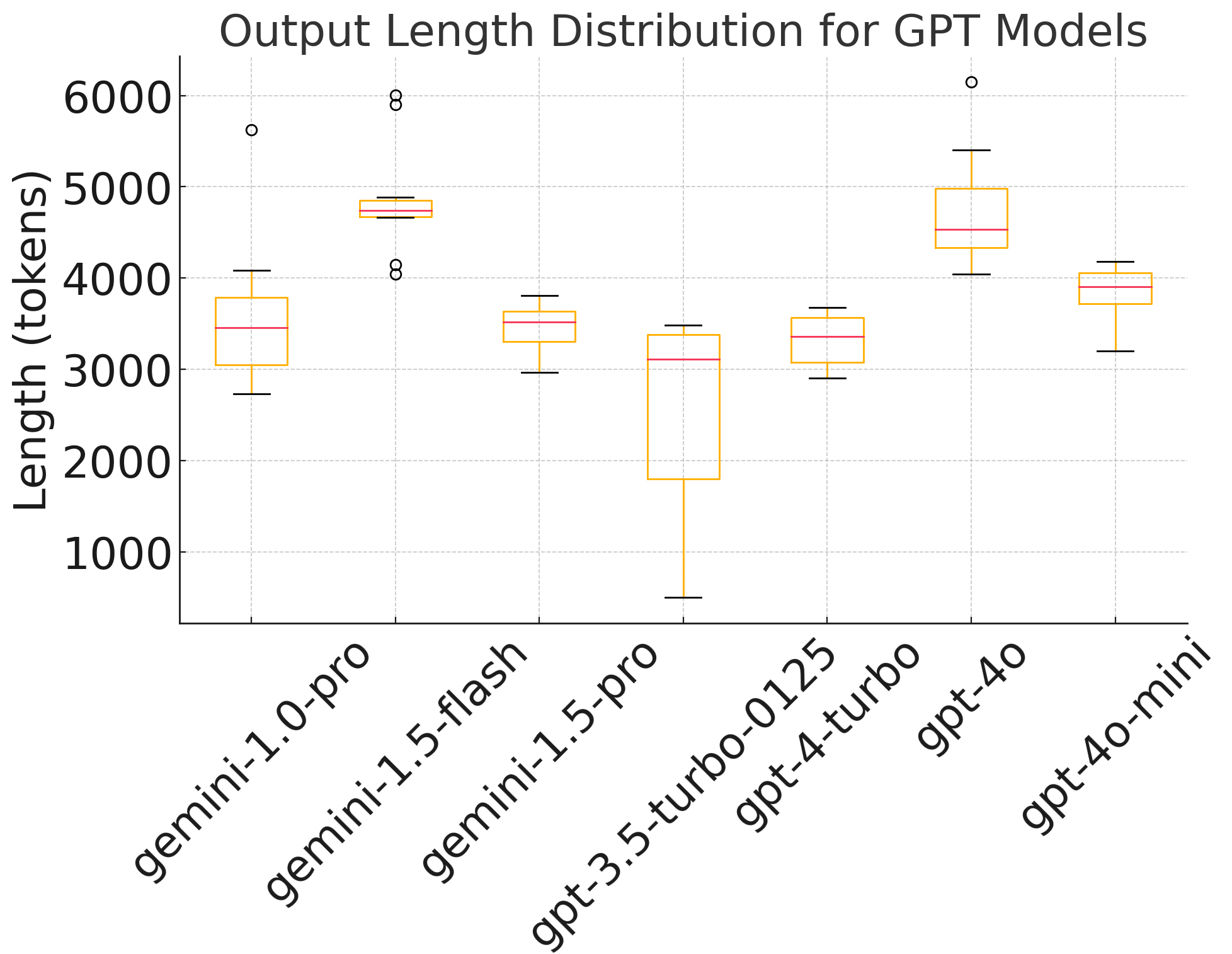}
		\caption{Output Tokens of different LLMs. (Stability of long text.)}
		\label{fig:OutputCloudComparison}
	\end{subfigure}
	\hfill
	\begin{subfigure}[b]{0.48\textwidth}
		\centering
		\includegraphics[width=\textwidth]{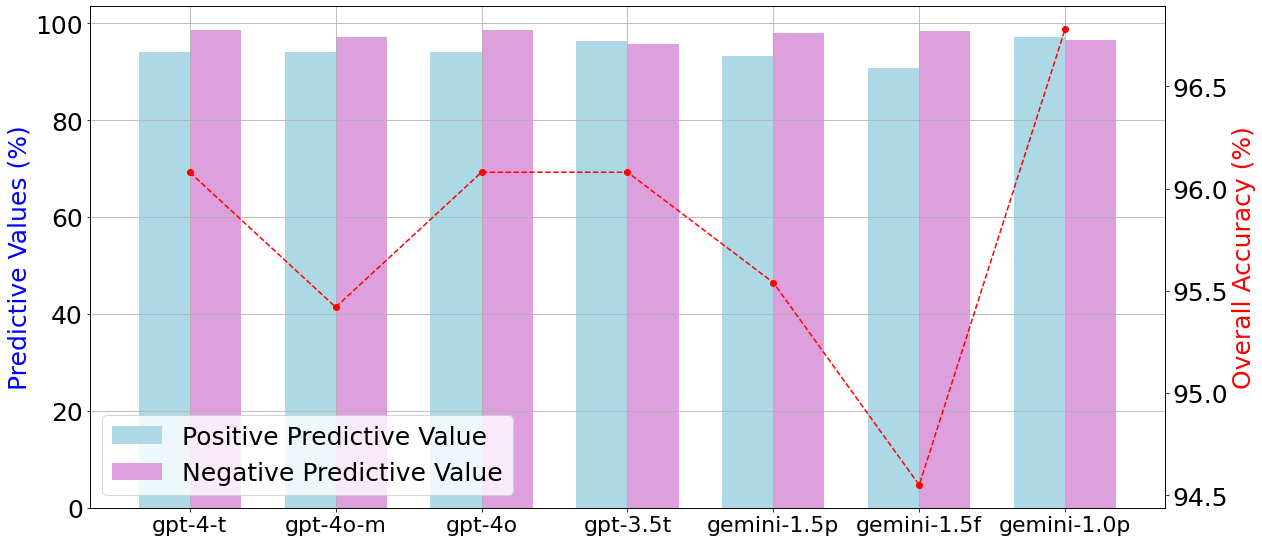}
		\caption{Sentiment analysis accuracy of LLMs.}
		\label{fig:AccurayDifferntModel}
	\end{subfigure}
	\hfill
	\begin{subfigure}[b]{0.5\textwidth}
		\centering
		\includegraphics[width=\textwidth]{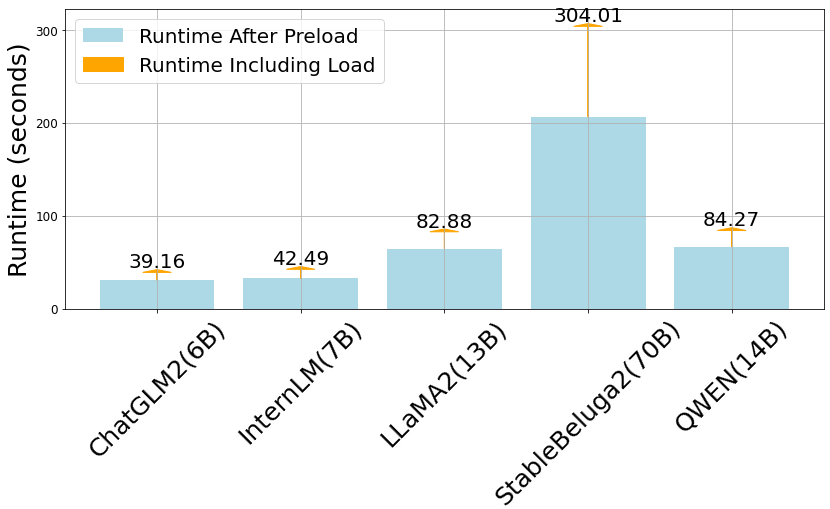}
		\caption{Run time on edge-level LLMs.}
		\label{fig:RunTimeLocalComparison}
	\end{subfigure}
	\caption{Performance of various large model deployment (cloud and edge).}
	\label{fig:test}
\end{figure}

\textbf{Cloud deployment} lies in leveraging the powerful long-form reasoning capabilities of models like \emph{ChatGPT} or \emph{Gemini}, ensuring stable and highly accurate inference while avoiding high hardware deployment costs. As illustrated in Fig.~\ref{fig:RunTimeCloudComparison} and Fig.~\ref{fig:OutputCloudComparison}, after retrieving local crawler documents, the LLM-based sentiment report can be completed in just 25 seconds. With an 4,000-words input, the performance of \emph{gpt-4o-mini} and \emph{gemini-1.5-flash} is already satisfactory and offers better cost-effectiveness compared to larger models in the same series (\emph{gpt-4-turbo} and \emph{gemini-1.5-pro}). 

\textbf{Sentiment accuracy} is invested by a dataset of more than 3,400 comments \cite{kaggle2024} and the sentiment analysis performance is validated  (positive or negative). Fig.~\ref{fig:AccurayDifferntModel} indicates that various commercial LLMs meet general sentiment analysis accuracy standards ($94\%$-$96\%$), although contradictions arise in cases where textual evaluations are positive yet receive low ratings.

\textbf{Edge deployment approach} utilizes lightweight large language models, enhancing reasoning capabilities for specific tasks while preserving privacy. Deployment of these large models requires only a simple PC or personal workstation. As shown in Fig.~\ref{fig:RunTimeLocalComparison}, we introduce several language models suitable for local deployment, including the continuously developed \emph{ChatGLM2, InternLM, LLaMA2, StableBeluga2}, and \emph{Qwend}. We monitor runtime performance under cold starts (including preload) and warm starts (after preload). For a near $4,000$-word input, the perplexity below $400$ is acceptable. Users prioritizing privacy protection select appropriately sized large models for local deployment based on SLA (service level agreement).

\subsection*{LLM based Sentiment Analysis}

In applying our collaborative AI system, we concentrated on analyzing the sentiment surrounding "takeout" across diverse media channels. Initially, we aggregated trending content from various sources, including \emph{Bing News}, \emph{Google News} (public media), \emph{Google Search} (search engines), and social media platforms like \emph{Twitter} and \emph{Yahoo Hot}. Sentiment analysis was then conducted using TextBlob. \textbf{TextBlob} is an open source text processing library that can be used to perform many natural language processing tasks, such as part-of-speech tagging, nominal component extraction, and sentiment analysis.

\begin{figure}
	\centering
	\begin{subfigure}[b]{0.4\textwidth}
		\centering
		\includegraphics[width=\textwidth]{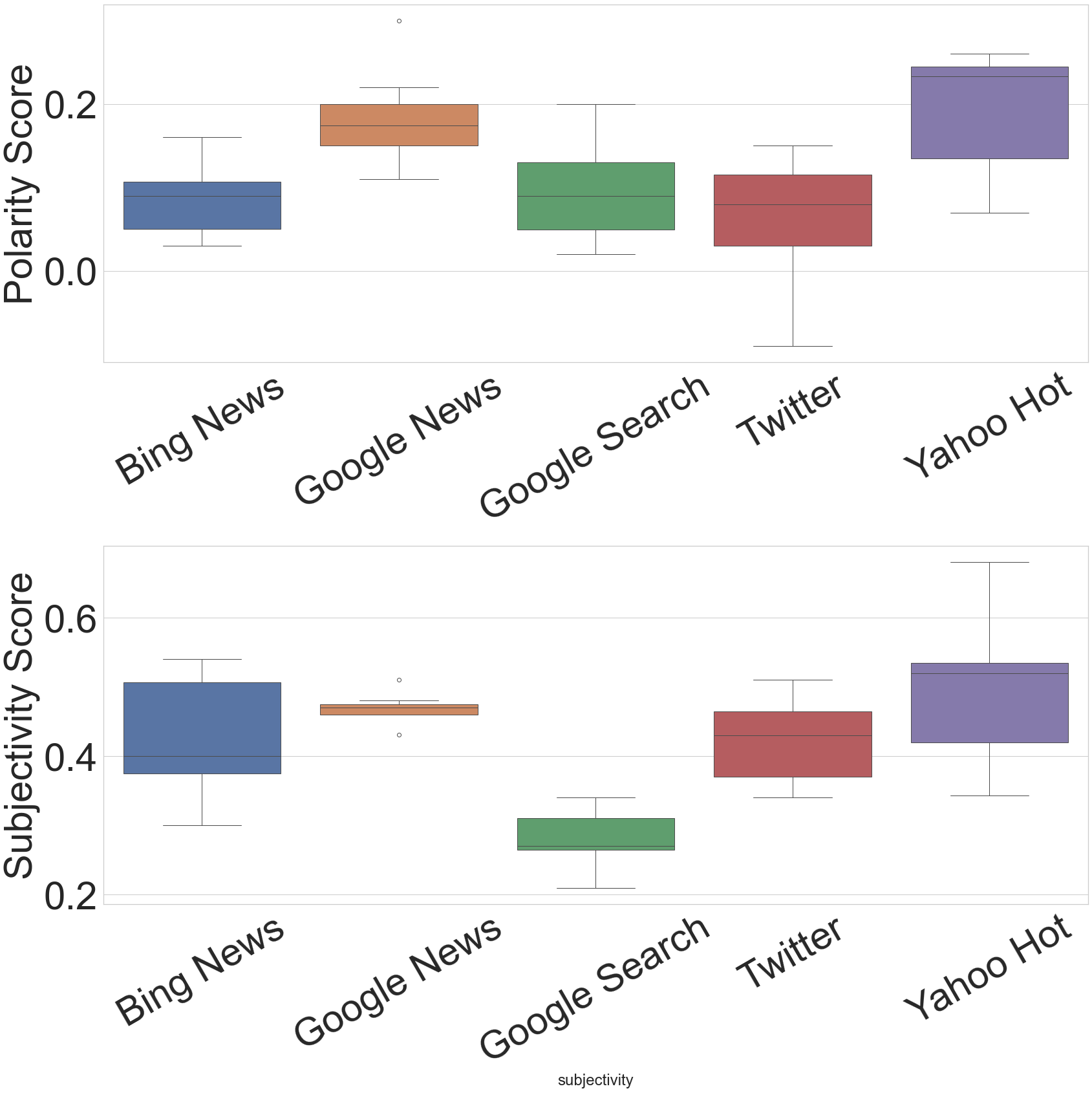}
		\caption{Polarity and subjectivity across different online media sources using stable MA.}
		\label{fig:sub1}
	\end{subfigure}
	\hfill
	\begin{subfigure}[b]{0.57\textwidth}
		\centering
		\includegraphics[width=\textwidth]{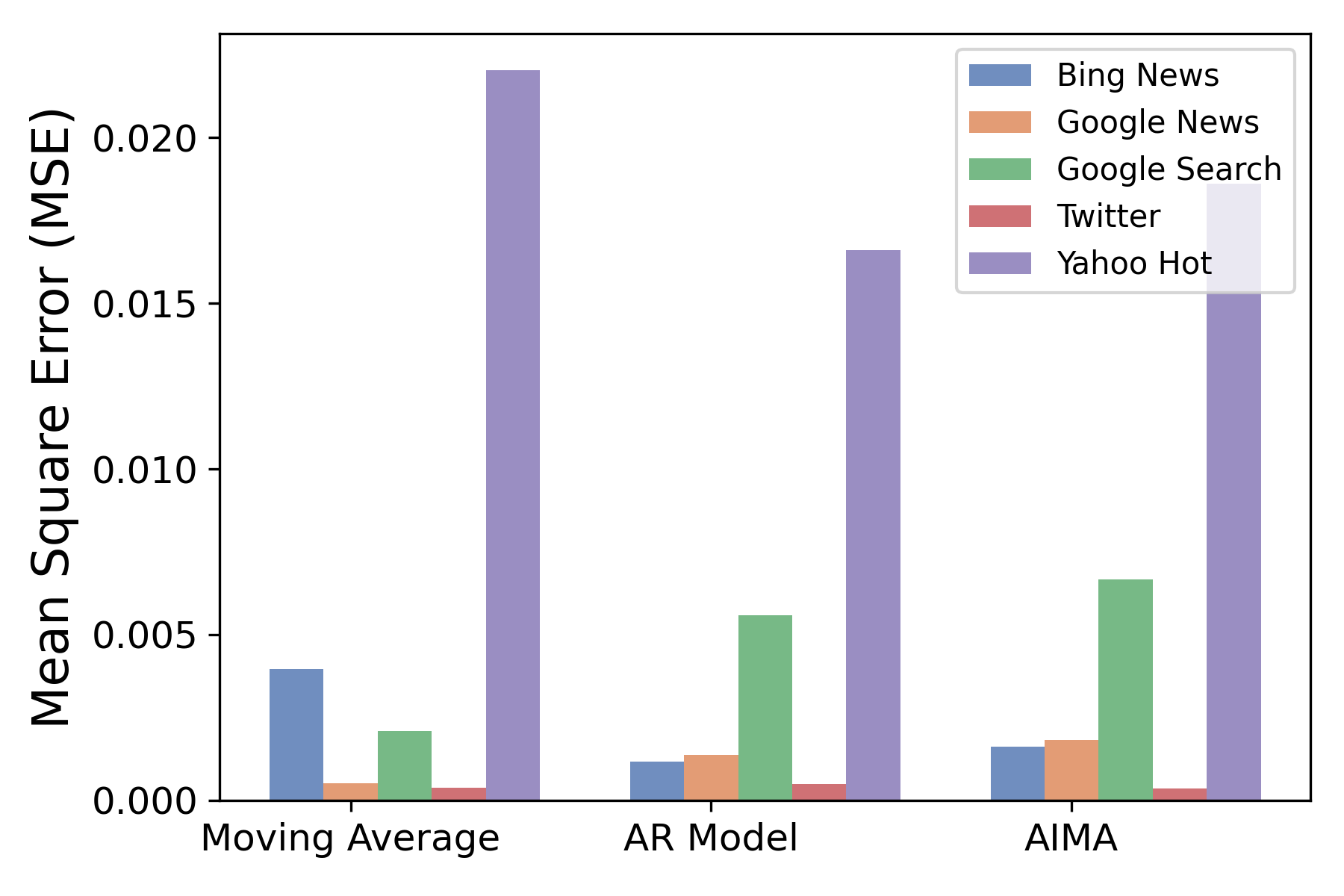}
		\caption{Diifferent prediction method comparison across typical online media sources.}
		\label{fig:sub2}
	\end{subfigure}
	\caption{Sentiment Analysis via popular online media sources.}
	\label{fig:test}
\end{figure}

Fig.~\ref{fig:sub1} illustrates our two-week sentiment analysis of "takeout". Notably, content from public media and search engines exhibited a more positive tone, in contrast to the significant variance seen on social media. Twitter, for instance, predominantly displayed negative sentiments, often reflecting user grievances, whereas Yahoo Hot tended towards positive sentiments, focusing on lifestyle sharing. These variations can be attributed to geographical diversity, lifestyle habits, platform-specific attributes, and content moderation policies, influencing the emotional resonance of topics within target demographics.
Objectively, "takeout" topics in public media maintain a professionally subjective narrative. Twitter, emphasizing problem feedback, demonstrates a rational tone, whereas Yahoo Hot, focusing on lifestyle sharing, exhibits a broader emotional range.

%Considering that LLM, as an emerging service, still requires extensive information collection, we make emotional predictions for the short-term mined information. Based on the five types of information channels mentioned, we use the Moving Average (MA), Auto Regression (AR) Model, and ARIMA (Autoregressive Integrated Moving Average) for regular predictions. 

As corroborated by Fig.~\ref{fig:sub2} , the stability of data variation significantly influences the predictive accuracy of the algorithms' emotional forecasts (Mean Square Error, MSE). We evaluate the discrepancy between the emotional true values (weighted average method) provided to users and the predicted values. The AR model demonstrates broad applicability in emotion predictions across various media channels. As the default algorithm for short-term data, its predictions are integral to the final report's content.

\section*{Conclusion}

With the increasing adoption of APIs like ChatGPT for basic tasks (e.g., after-sales service and FAQs) in businesses and online platforms, exploring their potential in professional fields and the substitutability of human and intellectual resources within organizations is of paramount interest. This paper presents a collaborative AI application paradigm, utilizing generative AI to deconstruct complex online sentiment analysis into interpretable, phased objectives. This includes defining the system, designing its architecture and operational mechanisms, and demonstrating implementation. After overcoming the challenge that LLM cannot handle multiple goals and complex logical tasks at one time, an automated solution including content search, inductive fusion, predictive analysis and other multiple goals was achieved, and the feasibility of cloud and edge deployment is verified. The system is expected to replace some human work in data collection, aggregation and analysis, and influence rational decision-making in corporate work through the sentiment analysis reports provided, thereby further improving the operating efficiency of corporate organizations.

\section*{ACKNOWLEDGMENTS}

This work is partially supported by JSPS KAKENHI Grant Numbers 22K17884, and 24K03055 Japan.  Chaofeng Zhang is the corresponding author.  E-mail: $ zhang-chaofeng@aiit.ac.jp$.

\newpage

\noindent{The following Cell Press journals require most research articles to include \textbf{experimental procedures} after the discussion: \textit{Chem}, \textit{Chem Catalysis}, \textit{Cell Reports Physical Science}, \textit{Cell Reports Sustainability}, \textit{Device}, \textit{Joule}, \textit{Matter}, \textit{Newton}, \textit{One Earth}, and \textit{Patterns}. For all other Cell Press journals, please \textbf{delete} the contents of this page from the template and refer instead to the \textbf{STAR Methods} section on page 8.}

\section*{STAR METHODS}

%%%  The STAR Methods should appear in your main 
%%%  manuscript file after the figure legends, main 
%%%  table(s) and table legend(s).

\subsection*{Key resources table}

\begin{table}[h]
	\centering
	\small % Adjusts the font size to small
	\begin{tabular}{|p{7cm}|p{7cm}|p{3cm}|}
		\hline
		\textbf{REAGENT or RESOURCE} & \textbf{SOURCE} & \textbf{IDENTIFIER} \\ \hline
		Deposited data &  &  \\ \hline
		Amazon Reviews for Sentiment Analysis & \url{www.kaggle.com/datasets/bittlingmayer/amazonreviews} &  \\ \hline
		Software and algorithms &  &  \\ \hline
		Cooperative AI Development Platform (original platform) & \url{https://agent.wondervoy.com/auth} &  \\ \hline
		Public Product \& Software Exhibition & \url{https://wondervoy.ai/images/img_case_2.jp} &  \\ \hline
		TextBlob: Simplified Text Processing & \url{https://textblob.readthedocs.io/en/dev/} &  \\ \hline
		Google Gemini Development Code & \url{https://ai.google.dev/gemini-api/docs} &  \\ \hline
		OpenAI Chatgpt-4 Development Code & \url{https://platform.openai.com/docs/overview} &  \\ \hline
	\end{tabular}
	\caption{Key resources table}
	\label{tab:key_resources}
\end{table}

\subsection*{Resource availability}

%%%  The resource availability section is required 
%%%  for all research articles. It is the first section 
%%%  that should appear under the STAR Methods heading 
%%%  (and will be typeset to appear directly below the 
%%%  key resources table). This section must contain 
%%%  the following required subsections: "lead contact," 
%%%  "materials availability," and "data and code 
%%%  availability." These 3 subsections are mandatory 
%%%  even if no unique reagents were generated in the 
%%%  study. Do not edit or change the names of the 
%%%  (sub)headings. No other subheadings or text are 
%%%  allowed in the resource availability section.

\subsubsection*{Lead contact}

%%%  Authors are required to designate a lead contact, 
%%%  who will be responsible for communication with 
%%%  the journal before and after publication and is 
%%%  the arbiter of disputes, including concerns 
%%%  related to reagents or resource sharing. Only 
%%%  one author can be named the lead contact, and 
%%%  only the lead contact’s information may be 
%%%  provided in this section.

Requests for further information and resources should be directed to and will be fulfilled by the lead contact, Chaofeng Zhang (zhang-chaofeng@aiit.ac.jp).

\subsubsection*{Materials availability}

%%%  This subsection must include a statement 
%%%  describing the availability of newly generated 
%%%  materials associated with the paper, including 
%%%  any conditions for access. If there are no 
%%%  newly generated materials associated with the 
%%%  paper, the statement should state this, e.g.: 
%%%  "This study did not generate new materials."

This study did not generate new unique physical products.

\subsubsection*{Data and code availability}

%%%  This subsection consists of three bullet points. 
%%%  Each bullet point is required. 

%%%  The first bullet describes availability of data. 
%%%  You must be willing to share the data reported 
%%%  in your paper after it is published, either 
%%%  through an online repository or upon request 
%%%  from the lead contact. 

%%%  The second bullet describes availability of 
%%%  custom code, either from a repository that 
%%%  provides DOIs (e.g., Zenodo) or as a supplemental 
%%%  item in the paper. If none was generated, please 
%%%  state “This paper does not report original code.”

%%%  The final bullet point must state: “Any 
%%%  additional information required to reanalyze 
%%%  the data reported in this work paper is 
%%%  available from the lead contact upon request.” 

The dataset and code to replicate the main analysis is deposited on software development platform (original) 
(\url{https://agent.wondervoy.com/auth})  and is publicly available as the SNS API   
(\url{https://wondervoy.ai/images/img_case_2.jpg}).

%%%  Please list here under separate headings 
%%%  all the experimental models/study participants 
%%%  (animals, human participants, plants, microbe 
%%%  strains, cell lines, primary cell cultures) 
%%%  used in the study. For each model, provide 
%%%  information related to their species/strain, 
%%%  genotype, age/developmental stage, sex (and 
%%%  gender, ancestry, race, and ethnicity if 
%%%  reported for human studies), maintenance, 
%%%  and care, including institutional permission 
%%%  and oversight information for the studies 
%%%  the experimental animal/human participant 
%%%  study. The influence (or association) of sex, 
%%%  gender, or both on the results of the study 
%%%  must be reported. In cases where it cannot, 
%%%  authors should discuss this as a limitation 
%%%  to their research’s generalizability.

%%%  Please omit this section if your study does 
%%%  not use experimental models typical in the 
%%%  life sciences (e.g., if your study is 
%%%  computational or physical science research). 

\subsection*{Method details}

\subsection*{Assessment and Prediction}
Comprehensive performance assessments of the collaborative AI system are integral to the final report generation. We employ mainstream, stable methodologies accessible via LLM's API for quantifiable emotion assessment and prediction.

\textbf{Assessment Algorithm:}
We integrat \textbf{TextBlob}, a versatile tool combining various NLP functionalities such as sentiment analysis, part-of-speech tagging, and noun phrase extraction to facilitate robust and universal emotion analysis. TextBlob's accessibility enables developers to implement it efficiently without needing extensive NLP expertise. Its computational efficiency and modular nature are particularly advantageous for handling large data volumes in collaborative AI systems. We utilize TextBlob to compute two key emotional indicators: {\emph{Polarity}} (ranging from $-1$ for negative to $1$ for positive sentiment) and {\emph{Subjectivity}} (from $0$ for objectivity to $1$ for subjectivity). Inspired by human semantic analysis \cite{habermas1990moral}, These metrics together establish a comprehensive sentiment analysis framework. Our methodology involves a weighted combination of these scores:
\begin{equation}
	\emph{Score} = w_{p} \cdot \emph{Polarity} + w_s \cdot \emph{Subjectivity}
\end{equation}
where  $w_p$ and $w_s$ represent the weights for Polarity and Subjectivity, respectively.

Adjusting these weights allows for tailored analysis relative to specific keywords, such as prioritizing \emph{Polarity} for consumer goods and \emph{Subjectivity} for tools and policies.
Other assessment tools are also viable options in specific application scenarios. For instance, NLTK (Natural Language Toolkit) offers a broader range of natural language processing capabilities, suitable for more complex text analysis tasks. Stanford NLP and spaCy are powerful NLP tools with differences in accuracy and performance, suitable for advanced language processing needs. Furthermore, Google Cloud Natural Language API, as a commercial service, provides large-scale and highly customizable text analysis capabilities. In designing collaborative AI systems, different tools help us more effectively achieve specific goals.

\textbf{Prediction Algorithms:}
When sufficient assessment data aggregates into a time series, we incorporate prediction algorithms to forecast future emotional trends. Long Short-Term Memory (LSTM), a quintessential AI technique within Recurrent Neural Networks (RNNs), is particularly adept at handling long sequence predictions. When the scale and complexity of the data reach significant levels, RNN-based networks are preferable for achieving precise forecasts \cite{chaudhri2021implementation}. Consequently, we utilize methods optimized for time series prediction with minimal data requirements:
\begin{itemize}
	\item \emph{Moving Average \textbf{(MA)}}: An elementary technique ideal for illustrating trends and smoothing data fluctuations.
	\item \emph{Auto Regression \textbf{(AR)} Model}: Suitable for capturing autocorrelation in datasets.
	\item \emph{Autoregressive Integrated Moving Average \textbf{(ARIMA)}}: A sophisticated model that merges autoregression and moving average attributes, effectively applied to non-seasonal data.
\end{itemize}
For these models, typically, only the daily sentiment scores from the past few days (such as average Polarity and Subjectivity) are necessary to compute relatively accurate predictions based on historical trends.

\end{document}